\documentclass[size12,A4paper]{article}
\usepackage[T1]{fontenc}
\usepackage{latexsym}
\usepackage{amssymb}
\usepackage{amsmath}
\usepackage{amsthm}
\usepackage{booktabs}
\usepackage{graphicx}
\usepackage{color}
\theoremstyle{definition}
\newtheorem{examp}{Example}[section]
\theoremstyle{remark}
\newtheorem{remark}{Remark} 
\usepackage{tikz}
\usepackage{tikz-qtree}
\usepackage{hyperref}
\hypersetup{
  colorlinks=true,
  urlcolor=blue,    
  linkcolor=black,  
  citecolor=blue,   
}
\UseRawInputEncoding

\title{A note  reviewing Turing's  1936}
\author{Paola Cattabriga\\
\small{University of Bologna - Italy} \\ \small{\href{https://orcid.org/0000-0001-5260-2677}{0000-0001-5260-2677} } }
\date{}
\begin{document}
\maketitle

\begin{abstract}
By closely rereading the original Turing's 1936 article, we can gain insight about that it  is based on the claim  to have defined a number which is not computable, arguing that there can be no machine computing the diagonal on the enumeration of the computable sequences.
This article provides a careful analysis of Turing's original argument, demonstrating that it cannot be regarded as a conclusive proof. Furthermore, it shows that there is no evidence supporting the existence of a defined number that is not computable.
\end{abstract}
\section{Introduction}
 As well known, Turing historical article of 1936  entitled  ``On Computable Numbers, with an Application to the Entscheidungsproblem'' is the result of a 
	special endeavor focused around the factuality  
	 of a general process for algorithmic computation. As  
	resultant formal model  his famous abstract computing machine, 
	soon called Turing  machine,
	could be regarded to be a universal feasibility 
	test for  computing procedures. 
	The article begins by accurately outlining the notion of a computable number; that is, a real number is computable only if there exists a Turing machine that writes all the sequence of its decimal extension. The abstract machine as a universal feasibility test for computing procedures is then applied up to closely examining what are considered to be the limits of computation itself and to defining a number which is not computable.
\begin{quote} The 
	computable numbers do not include, however, all definable numbers; and 
	an example is given of a definable number which is not 
	computable (230 \cite{turing36}).\end{quote} 
	In Section 8. is reached  the crucial demonstration 
	establishing some fundamental limits of computation by defining 
	such a number through a self-referring procedure. The present
	note  shows how this procedure can not  actually be regarded 
	as a demonstration.

	There is considerable literature introductory to Turing machines; see just a few  \cite{cop,davis1,hermes,kleene,trak,yasuhara} and also \cite{davis2,davis3,salvador,zao}. In the following, the reader is required to know the Turing article together with the original notions and symbolism therein contained 	\cite{turing36}. We will briefly remind the reader of just a few of the main ones, with some examples. Notions not mentioned in this introduction will be dealt  step by step with direct references to the pages of Turing  \cite{turing36},  hereinafter referred to as T36.

	\noindent
	\begin{trivlist}
	 \item [] {\it Computing machines.}  If any automatic machine $\mathcal M$ prints two 
	kinds of symbols, of which the first kind consists entirely 
	of $ 0 $ and $ 1 $  (the others being called symbols of the 
	second kind), then the machine will be called a computing 
	machine. If the machine is supplied with a blank tape and set 
	in motion, starting from the correct initial configuration, 
	the subsequence of the symbols printed by it which are of the 
	first kind will be called the {\it sequence computed by the machine}.
	\medskip
	 \item [] {\it Complete configuration.} 
	The real number whose expression as a binary decimal is obtained by prefacing 
	this sequence by a decimal point is called the number computed by the machine. 
	At any stage of the motion of the machine, the number of the scanned square, 
	the complete sequence of all symbols on the tape, and the $m$-configuration will be 
	said to describe the complete configuration at that stage. The changes of the machine 
	and tape between successive complete configurations will be called the moves of the machine.
	\medskip
	\item[] {\it Circular and circle-free machines. } If a computing 
	machine $\mathcal M$  never writes down more than a finite number of 
	symbols of the first kind, it will be called {\it circular}. 
	Otherwise it is said to be {\it circle-free}. 
	A machine will be circular if it reaches a configuration from 
	which there is no possible move, or if it goes on moving, and 
	possibly printing symbols of the second kind, but cannot 
	print any more symbols of the first kind.
	\medskip
	\item[] {\it Computable sequences.} A sequence is said to be 
	computable if it can be computed by a circle-free machine.
	\medskip
       \item[] {\it Computable numbers.} A number is computable if it 
       differs by an integer from the number computed by a 
       circle-free machine.
       \medskip
       \item[] {\it S.D. } Any  automatic machine $\mathcal M$ is identified by its Table 
       describing configurations and behaviors. Any Table  can be 
       coded or rewritten in a new description called the {\it  
       Standard Description} of $\mathcal M$ (Example \ref{tb_fig}).
       \end{trivlist}
       
  \medskip     
       
\begin{examp}\label{tb_fig}  
The table of $m$-configurations of a machine $\mathcal{ M}$ computing the infinite sequence  $01010101 \dots$   			
\[
\begin{array}[c]{llll}
q_1 & S_0 & PS_1 , R& q_2  \\
q_2 & S_0 & PS_0 , R& q_3\\
q_3 & S_0 & PS_2 , R& q_4\\
q_1 & S_0 & PS_0 , R& q_1\\
 \end{array}
 \]  
 which can be arranged on a line
   $$ q_1  S_0  PS_1 , R  q_2 ; \;
q_2  S_0   PS_0 , R  q_3 ; \;
q_3   S_0   PS_2 , R  q_4 ; \;
q_1   S_0   PS_0 , R  q_1;  . $$

\medskip

\noindent Standard Description  of $\mathcal{ M}$ 

\medskip

\noindent
{ \scriptsize DADDCRDAA ; \;
DAADDRDAAA ; \;
DAAADDCCRDAAAA ; 
DAAAADDRDA ; }

\medskip

\noindent Description Number of $\mathcal{ M}$ 
$$31332531173113353111731113322531111731111335317$$
\end{examp}

\medskip

\begin{trivlist}
	 \item[] {\it D.N.} Any letter in the standard description of $\mathcal M$ can be 
	 replaced by a number, so we shall have a description of the 
	 machine in the form of an  arabic numeral. The integer 
	 represented by this numeral is called {\it Description 
	 Number.} A number which is a description number of a 
	 circle-free machine will be called a {\it satisfactory} number (Example \ref{tb_fig}).
	       \medskip
	  \item[] {\it Universal Machine.} A universal machine is a computing machine  $\mathcal U$ 
	  that,  supplied with a tape on the beginning of which is 
	  written the {\it S.D.} of a computing machine $\mathcal 
	  M$,   computes the same sequence of $\mathcal M$.
	  \end{trivlist}
	  
\medskip
	  
	  A simple representation to view the Universal Machine in modern terms in Figure~\ref{uuuu}  on
page~\pageref{uuuu}.
	  \begin{figure}
 \begin{center}	  
	   input $ $ \underline{{\it S.D.}  of $\mathcal 
	  M$} $\longrightarrow $  $\:\: \boxed {\mathcal  U }\:\: $ $\longrightarrow $   output $ $  \underline {sequence computed by $\mathcal M$}	  
    \end{center}
    \caption{A representation of $\mathcal U$}\label{uuuu}
	\end{figure}
	We remark on the distinction between $m$-configuration   and   {\it complete configuration}  in T36.  A table of $m$-configurations is as in  previous Example \ref{tb_fig} (machine I  \cite[233]{turing36}). 
	An example of a list of the succesive {\it complete configurations} is the table of the sequence of symbols printed on the tape, by a machine during its computation, as arranged in the list ($\bf{C}$)  (machine II \cite[235]{turing36}):
	\begin{equation}\tag{\bf{C}}
	\mathfrak{b} : \; ^{\rotatebox{180}{e}} \;  ^{\rotatebox{180}{e}}  \; \mathfrak{o} \;  0 \quad
	0 : \; ^{\rotatebox{180}{e}} \;  ^{\rotatebox{180}{e}}  \; \mathfrak{q} \;   0 \quad 
	0 :  \; ^{\rotatebox{180}{e}} \;  ^{\rotatebox{180}{e}}  \; 0 \quad \; \mathfrak{q} \;
	0 : \;  ^{\rotatebox{180}{e}} \;  ^{\rotatebox{180}{e}}  \; 0 \quad \; 0 \; \mathfrak{q} \; 
	:  \;  ^{\rotatebox{180}{e}} \;  ^{\rotatebox{180}{e}}  \; 0  \quad 0 \; \mathfrak{p} \;  1 
	: \dots 
	\end{equation}
	A sequence of the complete configurations of a circle-free machine is an infinite table, i.e. an infinite sequence of symbols on the tape.  A sequence of the complete configurations of a circular machine is a finite sequence of symbols on the tape.

	The infinite sequence of complete configurations ($\bf{C}$) can then be coded in its standard form {\it S.D.} as in 
	\begin{equation}\tag{$\bf{C_1}$}
{\scriptsize DA : \,
 DDCCCDCCCDAADCDDC : \,
DCCCDCCCDAAADCDDC : \, \dots
 }
	\end{equation}
	and	
	\begin{equation}\tag{$\bf{C_2}$}
	{\scriptsize DDA : 0: 0: 
 DDCCCDCCCDAADCDDC : 
DCCC \, \dots
 }
	\end{equation}
	(machine II \cite[242]{turing36}).  Precisely, ($\bf{C_2}$) is ($\bf{C_1}$)  printing the figures of $\mathcal M$ on the tape. 
	Therefore, standard descriptions encode both the infinite table of successive complete configurations of a circle-free machine and  the finite table of the $m$-configurations of any machine. Although the terms ``$m$-configuration''
and successive ``complete configurations''  never appear in the later Section 8 on diagonalisation, they are fundamental concepts for understanding the entire subject matter contained therein \cite[246]{turing36}.  In the case of circle-free machines, it is certainly worth noting that  the infinite sequence in a format like ($\bf{C_2}$)  is  in a one-to-one correspondence with the complete sequence of figures $\mathcal M$ prints on the tape.
By convention we symbolize the computable sequence of $\mathcal  M$ with $C.S.(\mathcal  M)$, referring to the printed figures of a table of format  ($\bf{C_2}$).  Descriptive numbers  for ($\bf{C_1}$) and ($\bf{C_2}$) can also be encoded, respectively the  {\it D.N.} of ($\bf{C_1}$) and the  {\it D.N.} of ($\bf{C_2}$).
Both $C.S.(\mathcal  M)$ and {\it D.N.} of the ($\bf{C_2}$)-shaped table for circle-free machines are infinite sequences. The {\it D.N.} of the ($\bf{C_2}$)-shaped table are arabic numerals, with the figures 0 and 1 added, printed on the tape between the machine's successive moves. The $C.S.(\mathcal  M)$ are binary sequences: the same sequence of 0 and 1 between the successive moves of the ($\bf{C_2}$)-shaped  table.
For any circle-free $\mathcal  M$, the ($\bf{C_2}$)-shaped table generates the sequence  $C.S.(\mathcal  M)$.

Such a distinction is also fundamental to the notion of Universal Machine. A standard  description of a machine $\mathcal  M$ is the coded table of its $m$-configuration, which is finite, while the standard description, {\it S.D.}, of its sequence of complete configurations can be an infinite sequence. Accordingly,  if $\mathcal  M$ is circle-free, its output is the infinite list of computable sequences of $\mathcal  M$, $C.S.(\mathcal  M)$, which is  for the  Universal Machine  $C.S.(\mathcal  U)$.  The {\it S.D.} of both  $\mathcal  M$ and $\mathcal  U$ are finite.


\medskip

We also note  in (Example \ref{tb_fig})  that the Standard Description 
		
	\begin{center}{ \scriptsize DADDCRDAA ; \;
DAADDRDAAA ; \;
DAAADDCCRDAAAA ; 
DAAAADDRDA ; }\end{center}

encodes without the final dot the table of $m$-configurations  that is however present in the first standard form:

   $$ q_1  S_0  PS_1 , R  q_2 ; \;
q_2  S_0   PS_0 , R  q_3 ; \;
q_3   S_0   PS_2 , R  q_4 ; \;
q_1   S_0   PS_0 , R  q_1;.   $$ \cite[241]{turing36}.
A few lines above, Turing states that  other tables could be obtained by adding irrelevant lines such as $ q_1  S_1  PS_1 , R  q_2 ; $. 	
	All this might bring to mind infinite tables of $m$-configurations.  However, it seems unlikely that this truly reflects Turing's  intentions. If it were a redundant instructions queue that the machine would never read, its only purpose would be to extend its descriptive number indefinitely. In this case, though, a function for eliminating redundant rules could be added; see for example \cite{compact}. We  also add that these machines first designed by Turing are, in today's terms, deterministic automata.
In a broader sense, any algorithm should be defined in every detail, and
	a deterministic machine with an infinite table of $m$-configurations cannot be considered a valid and sound algorithm  \cite{hermes}. 
\section{The diagonal process}\label{diag_process}
	At the beginning of Section 8. {\it Application of the 
	diagonal process.},  Turing intends to submit to his 
	machine's feasibility test the application of Cantor's 
	  non-denumerability of real numbers to the computable 
	sequences. He verifies  if the diagonal process is suitable  to 
	show also the non-denumerability of computable sequences.
		\begin{quote} It might, for instance, be thought that the 
	    limit of a sequence of computable numbers must be 
	    computable. This is clearly only true if the sequence of 
	    computable numbers is defined by some rule (246 
	    \cite{turing36}).\end{quote} 
	A brief and elegant diagonalization is then proposed as follows:
	
	\medskip
	
   \begin{align}	
	a_{1} & = \quad \phi _{1} (1)  \quad  \phi _{1}(2)    \quad 
	\phi_{1}(3)   	\quad \quad  \ldots \notag   \\
	& \quad\quad\quad\quad \ddots \notag \\
	a_{2}¥ & =  \quad  \phi_{2}(1)    \quad  \phi_{2}(2)   \quad \phi_{2}(3)    \quad \quad \ldots   \notag \\
	& \quad\quad\quad\quad\quad\quad\qquad  \ddots \notag \\
	a_{3}¥ & =   \quad \phi _{3}(1)  \quad   \phi _{3}(2)  \quad  \phi _{3}(3)   \quad  \quad  \ldots  \notag   \\
        & \vdots \quad\quad\quad\quad\qquad\qquad\qquad\qquad  \ddots \notag \\
	a_{n}¥ & =   \quad \phi _{n}(1)   \quad \phi _{n}(2)  \quad \phi _{n}(3)  \quad \quad \ldots \phi _{n}(n) \notag \\
	& \vdots \quad\quad\quad\quad\qquad\qquad\qquad\qquad \qquad\qquad\qquad \ddots \notag 
\end{align}

\medskip

\noindent
where $a_{n}$ are the computable sequences with the 
	figures $ \phi _{n}(m)$  (on to $0,1$), and $\beta $ is the 
	sequence with $1 - \phi _{n}(n) $ as its n-th figure.
	\begin{quotation}
	\begin{trivlist}
	 \item [] {$\; \; \star$) } 
	\; Since  $\beta$ is computable, there exist a number $K$  such that  
	$1-\phi _{n}(n) = \phi _{K}(n)  $ all $n$. Putting $ n  = 
	 K$, we have $ 1  =  2 \phi _{K}(K) $, i.e. $1$ is 
	even. The computable sequences are therefore not 
	 enumerable.  
	 \end{trivlist}
\end{quotation}

 Turing himself considers argument  $\star$)  fallacious as it presupposes the computability of $\beta $, which in turn presupposes the enumerability of computable sequences by finite means. For Turing, the problem of enumerating computable sequences would be equivalent to finding out whether a given number is the D.N. of a circle-free machine, and he seems certain that the feasibility test provided by his machine will show the impossibility of any such process.
The most direct proof of   impossibility  could  be to 
	show that a machine exists  that computes $\beta $.  
	Turing seems here to attribute to the reader a special 
	undefined incertitude, a feeling that ``there must be 
	something wrong''.  We will not dwell upon whether it should 
	be the reader or	Turing himself to have such 
	inconvenient or inexplicable feelings\footnote{
	A display of the inferential steps in $\star$), which is a direct application of Cantor's diagonalization, 
	offers perhaps some explicative insight about why it  is considered  fallacious here.
	\begin{align*}	
	\beta & =  1 - \phi _{1} (1)  \;\;  1 - \phi _{2}(2)     \;\;  1 - \phi_{3}(3)   	 \;\;   \ldots \; 1 - \phi_{n}(n)  \;\; \ldots \notag   \\
	 1 - \phi _{n}(n) = &\phi _{K}(n) \,\;\downarrow \qquad\qquad\quad\;\downarrow   \qquad\qquad\;\;\downarrow  \quad \quad\;\; \qquad\quad\downarrow \notag \\
	 K & =  \quad  \phi_{K}(1)    \qquad\quad \phi_{K}(2)    \quad\quad \; \phi_{K}(3)   \;\,  \ldots  \; \quad \phi_{K}(n) \;\; \ldots  \notag \\
	K = &n  \quad \;\; \quad\: \downarrow \qquad\qquad\quad\; \; \downarrow   \qquad\qquad\; \; \downarrow  \quad \quad\;\; \qquad \quad	  \downarrow \notag \\
	 n & =  \quad \, \phi_{n}(1)    \qquad\quad\; \;  \phi_{n}(2)    \qquad \;\, \phi_{n}(3) \; \ldots \quad\; \, \phi_{n}(n) \; \; \ldots  \notag \\
         \end{align*}
	}.
	So he chooses to test the 
	feasibility of such a general process for finding whether a 
	given number is the {\it D.N.} of a circle-free machine through a self-referring argument. His argumentation will not be 
	based on  $\beta $ but on constructing $\beta ^{\prime} $, 
	whose  n-th figure is 
	$\phi _{n}(n)$ i.e., the same diagonal sequence $\phi_{1}(1) 
	\phi_{2}(2)\phi_{3}(3)$ $\ldots \phi_{n}(n)\ldots$ .

\begin{figure}[h]
	    
	    $\mathcal H$
$
\begin{array}{llllllll}	
&R (1)&   \notag \\
&&\diagdown \notag \\
 &R (1) & R (2) & \notag \\
& &  & \diagdown \notag \\
&R (1)  &  R (2)   &  R (3)  \notag\\
& \quad \vdots &   \quad \vdots   & \quad \vdots  & \diagdown  \notag\\
& \quad \vdots &   \quad \vdots   & \quad \vdots  &  \quad \vdots \;  \diagdown \notag\\
& \quad \vdots &  \quad \vdots    & \quad \vdots &  \quad \vdots& R(N - 1)  \notag\\
& \quad \vdots &  \quad \vdots    & \quad \vdots &  \quad \vdots & \quad\vdots \; R(N)  \notag\\
& \quad \vdots &  \quad \vdots    & \quad \vdots & \quad \vdots  &\quad \vdots \quad \vdots \;  R(N + 1)  \notag\\
& \quad \vdots &    \quad \vdots  & \quad \vdots & \quad \vdots  & \quad \vdots \quad \vdots  \qquad \; \; \;\ddots  \notag\\
 \end{array} 
$

\begin{center}
$\qquad \qquad\qquad$ \begin{tikzpicture}[level distance=120pt, sibling distance=30pt,every node/.style = { align=right}]
\node {$N$}
         [grow=right]
         child {node {u $\rightarrow$ $R(N)= R(N-1)$}}
         child {node {s $\rightarrow$ $R(N)= 1 + R(N-1)$}};
  \end{tikzpicture}
\end{center}
\caption{a representation of $\mathcal H$ computing the diagonal  $\beta ^{\prime} $}\label{r_diag}
\bigskip \medskip
\end{figure}

\bigskip

\section{The main argument}\label{diago}
	The whole 
	section 8 is based on the ``proof'' that there cannot exist an 
	effective process constructing  $\beta ^{\prime} $, namely 
	there is no feasible process generating  $\phi_{1}(1) 
	\phi_{2}(2)\phi_{3}(3)\ldots \phi_{n}(n)\ldots$. 
	
	\bigskip
	
	Turing's ``proof''
	is by reductio ad absurdum, assuming 
	that such a process exists for real.      That would be, we have a 
	machine $\mathcal D$ that given the {\it S.D.} of any machine 
	$\mathcal M$ will test 
	if $\mathcal M$ is circular, marking the {\it S.D.} with ``$u$'', or is 
	circle-free, marking the  {\it S.D.}  with ``$s$'' (Figure~\ref{dddd}  on
page~\pageref{dddd}).

	\begin{figure}
	 \begin{trivlist}
	\item \qquad   input \underline{{\it S.D.}  of {$\mathcal M$} } $\longrightarrow $  $\:\: \boxed {\mathcal  D }\:\: $ $\longrightarrow $   output 
	\item $ $ 
	\item  $     \qquad \qquad \qquad \qquad   \quad  $ either \underline{ $\mathcal M$ is circular then  mark  {\it S.D.} with ``$u$''} 
	\item $ $ 
              \item  \quad \qquad \qquad \qquad    \qquad  \quad \qquad \qquad  \quad or
                \underline { $\mathcal M$ is not circular then mark {\it S.D.} with ``$s$''} 
	 \end{trivlist}
	\caption{A representation of   $\mathcal  D$ }\label{dddd}
 \medskip
		\end{figure}
	\begin{remark}\label{thepoint}
	We notice that in the assumption  about ${\mathcal  D }$, it is passed over in silence  how ${\mathcal  D }$ will actually verify that {\it S.D.} is $s$, since in this case ${\mathcal  D }$ would have to produce the infinite sequence of complete configurations as in  ($\bf{C_2}$)-shaped table and only then issue a verdict. 
	In T36, it is instead assumed that ${\mathcal  D }$  is simply able to issue a verdict  whether  {\it S.D.} is $s$ or $u$, since otherwise it would never reach a decision in the case of $s$.
	This is explicitly confirmed by Turing a few paragraphs later in the sentence ``For, \emph{by our assumption} about ${\mathcal  D }$, the decision as to whether $N$ is satisfactory
is reached in a finite number of steps '', \cite[247]{turing36}.
	\end{remark}

	The hypothesis of the argument is continued in Section 8 of T36  by 
	constructing a machine $\mathcal H$ by combining $\mathcal D$ 
	and $\mathcal U$. The machine $\mathcal U$  
	simulates  $\mathcal M$  and generates the computable sequence $\beta ^{\prime}$ (Figure~\ref{hhhh}  on
page~\pageref{hhhh}).

	\begin{figure}
	 \begin{trivlist}
	
	 \item  input \underline{{\it S.D.}  of {$\mathcal M$} } $\longrightarrow $  $\:\: \boxed {\mathcal  D }\:\: $ $\longrightarrow $   output  
	 \item $ $ 
	   \item $\quad\quad \quad\qquad\qquad $ either \underline{ $\mathcal M$ is circular then  mark  {\it S.D.} with ``$u$"} 
	   \item $ $ 
           \item  $ \quad \qquad \qquad \qquad \qquad\qquad\qquad$ or  \underline { $\mathcal M$ is not circular then mark {\it S.D.} with ``$s$"} and 
           \item $ $ 
            \item $ \quad  \qquad  \qquad \qquad  \qquad $  input \underline{{\it S.D.}  of {$\mathcal M$}} $\longrightarrow $  \, \boxed {\mathcal  U }\, $\longrightarrow $ 
            output \underline {computable sequence $\mathcal M$}
        
	  	 \end{trivlist}
	
	\caption{A representation of   $\mathcal  H$ }\label{hhhh}
 \medskip
		\end{figure}
	\bigskip

	  \noindent The machine   $\mathcal H$ would have its motion divided into 
	   sections as follows. In the first $N - 1$ sections, among other 
	   things, the integers $ 1,2, \ldots N - 1 $ will have been 
	   written down and tested by the machine  $\mathcal D$. A 
	   certain number, say $R(N - 1)$, of them will have been found to 
	   be the 
	    {\it D.N.}'s of circle-free machines. 
	    In the $N$-th 
	    section the machine $\mathcal D$ tests the number $N$.
	    If $N$ is satisfactory, {\sl i.e.}, if it is the 
	    {\it D.N.} of a circle-free machine, then $R(N) = 1 + R(N 
	    - 1)$ and the first $R(N)$ figures of the sequence of 
	    which a   {\it D.N.}  is $N$ are calculated. The 
	    $R(N)$-th figure of this sequence is written down as one 
	    of the figures of the sequence $\beta ^{\prime} $ 
	    computed by $\mathcal H$. If $N$ is not satisfactory, then $R(N) =  R(N 
	    - 1)$ and the machine goes on to the $(N + 1)$-th section 
	    of its motion (247 \cite{turing36}) (Figure~\ref{r_diag}  on
page~\pageref{r_diag}).
  
  So that $\mathcal H$ is such  that 
	 \begin{equation}\label{turingH} 
	 R(N)
	     \begin{cases} N= s & \text{then} \qquad 1+ R(N-1) \\
		 N = u  &\text{then} \qquad R(N-1)
		 \end{cases}
	 \end{equation}	 
	 \medskip
	 where the number $R(N)$ is the $R(N)$-th figure of  $\beta ^{\prime} $, 
	 generated by $\mathcal H$ (i.e. the $\phi_{n}(n)$-th figure).
  \medskip
  
 The whole argument leads apparently to contradiction when $\mathcal H$ encounters 
	itself, namely its own {\it D.N.} $K$, turning out to 
	be $\mathcal H$ in the meantime circular and circle-free (Figure~\ref{camarra}  on
page~\pageref{camarra}).  The computation of the first $R(K)-1$ figures would be carried out all right, but the instructions for calculating the $R(K)$-th would amount to ``calculate the first $R(K)$ figures computed by $\mathcal H$ and write down the $R(K)$-th''. This 
$R(K)$-th figure would never be found (see $R(K)$ in Figure~\ref{camarra}  on
page~\pageref{camarra}). An explanation already emerge considering Remark \ref{thepoint}, see next section.

\begin{figure}[h]
$\mathcal H$
$
\begin{array}{llllllll}	

& \quad \vdots &   \quad \vdots   & \quad \vdots  &  \quad \vdots \;  \diagdown \notag\\
& \quad \vdots &  \quad \vdots    & \quad \vdots &  \quad \vdots& R(N - 1) &  \,  R(K- 1) \notag\\
& \quad \vdots &  \quad \vdots    & \quad \vdots &  \quad \vdots & \quad\vdots \; R(N) &  \, R(K)  \,  would \: never \: be \: found \notag\\
& \quad \vdots &  \quad \vdots    & \quad \vdots & \quad \vdots  &\quad \vdots \quad \vdots \;  R(N + 1)  \notag\\
& \quad \vdots &    \quad \vdots  & \quad \vdots & \quad \vdots  & \quad \vdots \quad \vdots  \qquad \; \; \;\ddots  \notag\\
 \end{array} 
$
\medskip
\begin{center}
$\qquad $\begin{tikzpicture}[level distance=100pt, sibling distance=26pt,every node/.style = { align=right}]
\node {$N$}
         [grow=right]
         child {node {u $\rightarrow$ $R(N)= R(N-1)$}}
         child {node {s $\rightarrow$ $R(N)= 1 + R(N-1)$}};
  \end{tikzpicture}
\end{center}

\begin{center}
\begin{tikzpicture}[level distance=130pt, sibling distance=26pt,every node/.style = { align=right}]
\node {$K$}
         [grow=right]
         child {node {s $\rightarrow$ $R(K)= $ calculate the first $R(K)$ figures \\ computed by $\mathcal H$ and write the  $R(K)$-th}}
         child {node {u $\rightarrow$ $\varnothing$ $\quad \quad \quad $ $\quad \quad\quad \quad \quad $ $\quad \quad \quad\quad  $ $\quad \quad \quad $ } };
  \end{tikzpicture}
\end{center}
\caption{$\mathcal H$ encounters its own {\it D.N.} $K$}\label{camarra}
\bigskip \medskip
\end{figure}	 
	\section{Reviewing Turing's argument}	
	The machine design in T36 is based on the factuality of a general algorithmic computation process, with the first seven sections focusing on this.
		 We call the meticulous design of the machines with their moves, the plan of the effective procedures of computation. Or, in short, the \emph{effective}. The hypothetical argument in the eighth section, however, is strongly influenced by Cantor's diagonalisation, with particular reference to Hobson's conception of definable numbers \cite{zao}. Referring back to Remark \ref{thepoint}, this leads to a hypothetical argument that is also  disconnected from the \emph{effective} and entirely abstract as respect to the factual computation of the machines. From this point onwards, we refer to this plan as the \emph{abstract}.
 
  Each $C.S.(\mathcal M)$  is produced by a ($\bf{C_2}$)-shaped table, which continuously outputs the sequence computed by  a circle-free machine $\mathcal M$ in binary format, owing to the specific design of the machines.  If {\it D.N.}$(\mathcal M)$ is $s$, then the corresponding {\it D.N.}($\bf{C_2}$) is infinite, and $C.S.(\mathcal M)$ is generated. However, in any case, {\it D.N.}$(\mathcal M)$ is finite.
If  {\it D.N.}$(\mathcal M)$ is $u$ then the corresponding  {\it D.N.}$(\bf{C_2})$ is finite, and there is no {\it C.S.}$(\mathcal M)$.
	To determine whether {\it D.N.}$(\mathcal M)$ is $s$ or $u$ for any machine $\mathcal M$, it is essential to have a machine that initially generates the complete sequence produced by $\mathcal M$. 
	If $\mathcal M$ stops printing the binary sequence after a finite number of moves as encoded in the corresponding table ($\bf{C_2}$), then the sequence is $u$, and the machine could also be designed to print $u$ at the end.
Instead, if the binary sequence printed by table ($\bf{C_2}$) never stops being produced, then the sequence is $s$. But it would certainly not be possible to design the machine to print $s$ at the end.
By the \emph{effective}, this applies to all
machines, due to their intrinsic design.
	The assumption within Section 8 that $\mathcal D$ is capable to decide for any $\mathcal M$ whether it is $s$ or $u$ is not merely hypothetical, but also \emph{abstract}.
	The two planes, \emph{abstract} and \emph{effective}, are in conflict as soon as the argument assumes the existence of $\mathcal D$.  The argument assumes toward contradiction the existence of $\mathcal D$, when by the \emph{effective} it is already well established  that $\mathcal D$ cannot actually exist.
	It turns out that $\mathcal D$ and the entire argument in Section 8 are isolated, or detached, from the plane of the \emph{effective}  construction of the machines.
	The whole argument  is based on the assumption that $\mathcal D$ arrives at its verdict in a finite number of steps, which isn't feasible in case $s$.

		\bigskip

Remaining on the same hypothetical level of argument in Section 8 of T36, it is possible to propose a way out, as follows.
$\mathcal H$ is associated with  its Table of $m$-configurations and therefore to its {\it S.D.}. Accordingly, the assumption of the existence of $\mathcal H$ also contains  that  its {\it D.N.} $K$ is  finite, very well coded and known;  whole Turing's formalism was built for this purpose.  We will now show how the entire reasoning neglects the consequences of all this at the level of the machines themselves, which is a crucial aspect to consider  in coherence with their construction.
	 Contrary to what all subsequent literature after T36  has presupposed, nothing really  prevents us from defining  the machine $\mathcal H$ such 
	that if it encounters its {\it D.N.} $K$ does not 
	upload it in the $R(N)$-th figure of $\beta ^{\prime} $. 
We can show how to effectively define  $\mathcal H  $ with the instructions  such that 
	\begin{equation} \label{newH}
	R(N)
	     \begin{cases}
		 N= K & \text{then} \qquad R(N-1) \\	 
		 N= s & \text{then} \qquad 1+ R(N-1) \\
		 N = u  &\text{then} \qquad R(N-1)
		 \end{cases}
	 \end{equation}	
	 where $ R(N)$ is the $R(N)$-th figure of  
	 $\beta ^{\prime} $ without $R(K)$.   
	  Actually, when $H $ in (\ref{newH}) is in the $N$-th section such that $N = K$, 
	 $\mathcal H $  goes on to the $(N + 1)$-th section of its 
	 motion. So $K$, as well as the $u$  numbers, is not included 
	 in the $R(N)$-th figure of  $\beta ^{\prime} $.  So what does $\mathcal H$ do in 
	 the $K$-th section?  Simply $\mathcal H $  goes on 
	 to the $(K+1)$-th section of $\mathcal H$, and its computation would 
	 be carried on (Figure~\ref{sestina}  on
page~\pageref{sestina}). 

	  \begin{figure}[h]
	 $\mathcal H$
$
\begin{array}{llllllll}	
&R (1)&   \notag \\
&&\diagdown \notag \\
 &R (1) & R (2) & \notag \\
& &  & \diagdown \notag \\
&R (1)  &  R (2)   &  R (3)  \notag\\
& \; \;  \vdots &   \; \;  \vdots   & \; \;  \vdots  & \diagdown  \notag\\
& \; \;  \vdots & \; \;  \vdots    & \; \;   \vdots &   \vdots \;\; R(N - 1)  \notag  \quad  R(K - 1)\\
&\; \;  \vdots &  \; \;  \vdots    &\; \;  \vdots &  \vdots  \quad  \vdots \;\;  R(N)  \notag  \quad   R(K)= R(K -1) \\
& \; \;   \vdots & \; \;  \vdots    & \; \;   \vdots &    \vdots  \quad \vdots \quad \vdots \;  R(N + 1)  \notag   \quad  R(K + 1)\\\
& \; \;  \vdots &    \; \;   \vdots  &\; \;   \vdots &    \vdots   \quad \vdots \quad \vdots  \qquad \; \;\ddots  \notag\\
 \end{array} 
$
\begin{center}
$\qquad \qquad \qquad \qquad \qquad$\begin{tikzpicture}[level distance=122pt, sibling distance=22pt,every node/.style = { align=right}]
\node {$N$}
         [grow=right]
         child {node {u $\rightarrow$ $R(N)= R(N-1)$}}
         child {node {s $\rightarrow$ $R(N)= 1 + R(N-1)$}}
         child {node {$N = K$ $\rightarrow$ $R(N)= R(N-1)$}};
  \end{tikzpicture}
\end{center}
\caption{ $\mathcal H$ keep computing when encouters $K$}\label{sestina}
\bigskip \medskip
\end{figure}

We cannot then state that $\mathcal H  $ in (\ref{newH}) is 
	 circular like $\mathcal H$ in (\ref{turingH}). 
	 When $K$ is encountered, $\mathcal H$ in (\ref{turingH}) stops, but  $\mathcal H $ as defined in (\ref{newH}) continues computing the computable sequence of $\mathcal H$. 
	   It is clearly possible to add a proviso at the beginning before the input {\it S.D.}($\mathcal M$), such that to exclude $K$ from the computation of  $\mathcal H$, see  Figure~\ref{hn}  on
page~\pageref{hn}. In simple terms, it is not required that $ \mathcal H$  computes its own {\it D.N.} $K$. 
Since  $K$ is the {\it D.N.}($\mathcal H$), it is finite and known from the beginning. That is, it is known as soon as the table of $m$-configurations of $\mathcal H$ is defined.

\begin{figure}
	 \begin{trivlist}
	 
	 \item \underline{{\it S.D.}  of {$\mathcal M$}} $\longrightarrow $ \underline{{\it D.N.} of {$\mathcal M$}}  and  \underline{{\it D.N.} of {$\mathcal M$}  $\neq$ {$K$}}
	  \item $ $ 
	 \item  input \underline{{\it S.D.}  of {$\mathcal M$} } $\longrightarrow $  $\:\: \boxed {\mathcal  D }\:\: $ $\longrightarrow $   output  
	 \item $ $ 
	   \item $\quad\quad \quad\qquad\qquad $ either \underline{ $\mathcal M$ is circular then  mark  {\it S.D.} with ``$u$"} 
	   \item $ $ 
           \item  $ \quad \qquad \qquad \qquad \qquad\qquad\qquad$ or  \underline { $\mathcal M$ is not circular then mark {\it S.D.} with ``$s$"} and 
           \item $ $ 
            \item $ \quad  \qquad  \qquad \qquad  \qquad $  input \underline{{\it S.D.}  of {$\mathcal M$}} $\longrightarrow $  \, \boxed {\mathcal  U }\, $\longrightarrow $ 
            output \underline {computable sequence $\mathcal M$}
        
	  	 \end{trivlist}
	
	\caption{A representation of   $\mathcal  H$ without selfreferring }\label{hn}

		\end{figure}
		
 $\mathcal H$ in (\ref{newH}) is accordingly always $\mathcal H$ in (\ref{turingH})  not computing itself. Remaining on the same \emph{abstract} plane as Turing's argument,  it seems that a Turing machine $\mathcal H $ without self-reference could be feasible, and there is no conclusion that there is no machine which computes $\beta^{\prime}$.
Just as at the \emph{effective} level the assumption of the existence of  $\mathcal D$ is nonsensical,   even at the \emph{abstract} level
we have no  conclusion that there can be no machine $\mathcal D$.

	It should be noted and emphasised  that the initial proviso in Figures~\ref{hn}  on
page~\pageref{hn} is based on the fact that  $K$ is a finite and known number. 
Simply, a number that has already been given is excluded. In this way, $\mathcal  H$ will not be computing  itself.
Indeed, there would be an explicit self-reference of $\mathcal  H$ in (\ref{newH}) if and only if the initial condition had been ``{\it S.D.}  of {$\mathcal M$} $\neq$ {\it S.D.} of {$\mathcal H$''. In this latter case, we could even arrive to a sequential hierarchy  of $\mathcal H$-machines. It could easily be shown that such a hierarchy would be still computable since  {\it S.D.}({$\mathcal M$}) and {\it S.D.}({$\mathcal H$) are finite. However this is not actually the case, since the proviso is directly  ``{\it D.N.} of {$\mathcal M$}  $\neq$ {$K$}'', and no hierarchy of  $\mathcal H$-machines is generated.

	So, still according to the hypothesis of T36's argument, we can always have machines not computing themselves, and  we don't reach the conclusion that there is no machine which computes  $\beta^{\prime}$.
	
\bigskip	
	 
\section{Definable numbers}
	Without taking the diagonal into consideration, let's just think about the sequence of $a_n$  as a simple list of all computable sequences,  as in section \ref{diag_process}.  Although the two-dimensional representation allows the diagonal to emerge as constructively generated step by step, the list of $a_n$'s is itself infinite.
	In $\phi_n(m)$, $m$ is the arabic numeral of the $m$-th figure of the binary computable sequence denumerated by  $n$. 
	The enumeration of all the computable sequences, the $n$-th $C.S.(\mathcal M)$, is $n$ in  $\phi_n(m)$, and $n$ is enumerably infinite.  And each $C.S.(\mathcal M)$ generates the sequence of the figures  $\phi_n(m)$, which is again infinite.
	 	
	\medskip
	
	 Now returning to T36's argument,  we have that $K$ is at the same time both  {\it D.N.}$(\mathcal H)$ and $a_{K} = C.S.(\mathcal H)$.  
	 However, this is impossible for $\mathcal H  $ in (\ref{turingH}), 
	 since $K$ as {\it D.N.}$(\mathcal H)$  is finite, whilst  $K$ in $a_K$,  as a computable sequence, is infinite.  For $\mathcal H  $ in (\ref{turingH}) the $R(K)$-th figure is $\phi_{K}(K)$, in which diagonalization is performed step by step, from $R(K-1)$ to $R(K)$.
	 However, 
	 from the point of view of the \emph{effective} computation, this neglects the fact that $K$ within T36's argument is two different numbers, since both are generated by two different machines: $K$ can be either a finite number (when it is {\it D.N.}$(\mathcal H)$) and an infinite number (when as {\it C.S.}$(\mathcal H)$ it is {\it D.N.}($\bf{C_2}$) of $(\mathcal H)$).
$K$ is two different numbers at the same
time, one finite and one infinite. Therefore, even if we state that the $R(K)$-th figure is $\phi_{K}(K)$  at  the \emph{abstract} level of diagonalization, we are neverthenless assuming a number that, at the level of  \emph{effective} computation, is impossible.

	Suspending judgement on how nonsensical the argument for $\mathcal H  $ in (\ref{turingH}) appears now: in the case of $\mathcal H$ as circle-free, $K$ as $s$ is both a finite and infinite number. 
	Let us distinguish in symbols  $K_{{\it D.N.}(\mathcal H)}$  from $K_{C.S.(\mathcal H)}$, where the first is $K$ as the {\it D.N.} of $\mathcal H$ and the second is the number of the computable sequence printed by  $\mathcal H$, i.e. $a_{K} = C.S.(\mathcal H)$.
	Accordingly, we realize that the entire argument assumption misrepresents them as if they were identical, because it merges the finite and the infinite.  Further as   $\mathcal H$ is assumed to compute  $\phi_{1}(1)  \phi_{2}(2)\phi_{3}(3)\ldots \phi_{n}(n)\ldots$, in the case of  $R(N)$ such that  $N = K$ and thus $R(K)$,  we have $$\phi_{K_{C.S.(\mathcal H)}}(K_{{\it D.N.}(\mathcal H)}).$$
	
	Since no number can be both finite and enumerably infinite at the same time, this is clearly a uniqueness violation in terms of the rules of the Theory of Definition. These rules exist to prevent superimpositions and circularity, and are based on the Criterion of Eliminability and the Criterion of Non-Creativity, originally developed by the Polish logician S. Le\v{s}niewski \cite{suppes,peruzzi,rogers}.   The point of introducing a new symbol, within a already well-established theory, is to facilitate deductive investigations from the structure of the theory, not to add to that structure. The rules established by the two Criterions state the conditions for proper equivalence and give some basic restrictions. To correctly define a new operational symbol, the uniqueness restriction should be applied, otherwise, a contradiction could occur \cite{catta,cattab,suppes,rogers,peruzzi}.
	Accordingly we can state  that
	\begin{equation}\label{tie}
	K_{D.N.(\mathcal H)}\;  \neq \; K_{C.S.(\mathcal H)}.
	\end{equation}
	For  more detailed discussions of  similar uniqueness violations due to  \emph{abstract}  forms of self-reference see \cite{catta,cattag,cattab}.
\medskip

	Furthermore, we could also argue that if $K$ is the {\it D.N.} of $\mathcal H$, then it is indeed  finite, and as a finite sequence of complete configurations, it should be the case that  $\mathcal H$ is circular. In accordance,  $K$ could not be the number of the computable  sequence  computed by $\mathcal H$ as circle-free. 	
	This now seems to be the true meaning of Turing's statement that $R(K)$ will never be found: since $\mathcal H$ is circle-free and $K$ is $s$, it should write as an infinite computable sequence  its own finite {\it D.N.}.
It is now clear that the uniqueness violation  (\ref{tie}) is due to the \emph{abstract} assumption concerning $\mathcal D$, according to which the decision whether $N$  is $s$ or $u$ is performed in a finite number of steps, which  \emph{effectively} cannot happen when $N$ is $s$  (compare with the last paragraph in (247, \cite{turing36})).  
\medskip

For all the reasons stated in the last two sections,  T36's so called  ``proof''  by reductio ad absurdum that it cannot exist an effective processing  constructing  $\beta^\prime$, fails to reach its conclusion.   We can then 
	 regard accordingly all the other arguments arising (248, 259-265 
	 \cite{turing36}).
	 Furthermore,  there is no evidence that $K$ is not effectively computable: although {\it D.N.}($\mathcal H$)  is by its construction theoretical,  $K$ is by its very definition the encoding {\it D.N.}($\mathcal H$). On account of (\ref{tie}), there is not even ``an example of a definable  number which is not computable''.  
	 
\medskip

	 Let us add that the notion of circle-free machines
	 echoes a lot the requirement that a definition must not  be  
	 circular, which is what  in the Theory of Definition is 
	 known to be ruled by the Criterion of Eliminability \cite{suppes,peruzzi}. A
	 definition that does not satisfy this requirement introduces, indeed,
	  a primitive term, and it is not a definition at all.
	 One might object that the construction of the number $K$ is not a definition, 
	 but this would not correspond to what is stated in T36.  
	 As shown in \cite{zao}, the concept of definable number adopted by Turing seems to be influenced by Hobson. However, to a modern reader, Hobson's conception may appear far from the Theory of Definition, which establishes the conditions for proper definitions giving some basic restrictions to prevent superimpositions and circularity, as originally by  Le\v{s}niewski  \cite{suppes,rogers,peruzzi,catta,cattab}. This, however,
should be the subject  of
further future investigations.

 \bigskip







\end{document}